\documentclass[preprint,aps]{revtex4}

\usepackage{amsmath}
\usepackage{amsfonts}
\usepackage{graphics}

\begin{document}

\newcommand{\unmedio}{{\scriptstyle\frac{1}{2}}}
\newcommand{\eff}{_{\text{eff}}}
\newcommand{\Infinity}{\infty}
\newcommand{\flip}{_{\text{flip}}}
\newcommand{\bos}{_{0\text{B}}}
\newcommand{\bosonic}{_{\text{bos}}}
\newcommand{\ferm}{_{0\text{F}}}
\newcommand{\trial}{_{\text{trial}}}
\newcommand{\true}{_{\text{true}}}
\newcommand{\Ci}{\operatorname{Ci}}
\newcommand{\tr}{\operatorname{tr}}
\newcommand{\PsiB}{\bar{\Psi}}   
\newcommand{\phiH}{\hat{\phi}}
\newcommand{\etaH}{\hat{\eta}}
\newcommand{\chiB}{\bar{\chi}}
\newcommand{\xiH}{\hat{\xi}}
\newcommand{\zetaH}{\hat{\zeta}}
\newcommand{\vH}{\hat{v}}
\newcommand{\bH}{\hat{b}}

\newcommand{\AB}{\bar{A}_\mu}
\newcommand{\BB}{\bar{B}_\mu}
\newcommand{\AT}{\tilde{A}_\mu}
\newcommand{\BT}{\tilde{B}_\mu}

\newcommand{\slp}{\raise.15ex\hbox{$/$}\kern-.57em\hbox{$\partial$}}
\newcommand{\slA}{\raise.15ex\hbox{$/$}\kern-.63em\hbox{$A$}}

\newcommand{\difp}{\frac{d^2p}{(2\pi)^2}\,}

\newcommand{\bra}{\left\langle}
\newcommand{\ket}{\right\rangle}
\newcommand{\bracket}{\left\langle\,\right\rangle}

\newcommand{\D}{\mathcal{D}}
\newcommand{\N}{\mathcal{N}}
\newcommand{\Lag}{\mathcal{L}}
\newcommand{\V}{\mathcal{V}}
\newcommand{\Z}{\mathcal{Z}}

\title{Improved harmonic approximation and the 2D Ising model at $T\neq T_{c}$ and $h\neq0$}
\author{An\'{\i}bal Iucci}
\author{Carlos M. Na\'on}
\email{iucci@fisica.unlp.edu.ar; naon@fisica.unlp.edu.ar}

\affiliation{Instituto de F\'\i sica La Plata. Departamento de F\'\i sica, Facultad de
Ciencias Exactas, Universidad Nacional de La Plata. CC 67, 1900 La Plata, Argentina.\\
Consejo Nacional de Investigaciones Cient\'\i ficas y T\'ecnicas, Argentina.}

\begin{abstract}
We propose a new method to determine the unknown parameter associated to a
self-consistent harmonic approximation. We check the validity of our technique in the
context of the sine-Gordon model. As a non trivial application we consider the scaling
regime of the 2D Ising model away from the critical point and in the presence of a
magnetic field $h$. We derive an expression that relates the approximate correlation
length $\xi$, $T-T_c$ and $h$.
\end{abstract}



\maketitle

The so called ``self consistent harmonic approximation" (SCHA) is a non-perturbative
technique that has been extensively employed in Statistical Mechanics
\cite{Saito}\cite{Fisher-Zwerger} and Condensed Matter physics
\cite{Gogolin}\cite{Egger}\cite{Xu}\cite{Iucci} applications. Roughly speaking it
amounts to replacing an exact action $S_{\text{true}}$ by a trial action
$S_{\text{trial}}$ that makes the problem tractable. Usually $S_{\text{trial}}$ is
just a quadratic action that depends on certain unknown parameter $\Omega$ that must
be determined through some criterion such as the minimization of the free energy of
the system. This approximation is intimately related the the ``gaussian effective
potential"  \cite{Stevenson}\cite{Ingermanson} in Quantum Field Theories (QFT's), a
variational approximation to the effective potential which uses a gaussian wave
functional depending on some mass parameter as the trial ground state. It also relies
on a minimization principle often called ``principle of minimal sensitivity"
\cite{minimal} to determine the additional parameter. In this work we point out that
in two-dimensional problems there is an alternative way to obtain the quantity
$\Omega$. This method is based on Conformal Field Theory (CFT) \cite{CFT}. Moreover,
we shall show that our method yields improved results with respect to the predictions
of standard SCHA in the sine-Gordon (SG) model and allows us to give a new description
of the off-critical 2D Ising model (2DIM). In the former we exploit the existence of
exact results \cite{DHN} \cite{Zamolodchikov2} to check the consistency of our
proposal by obtaining a qualitatively good answer for the soliton mass. We then apply
the same idea to the 2D Ising model at $T\neq T_c$ and $h\neq0$, a non-integrable
model in which very few quantitative results are known \cite{McCoy} \cite{Delfino et
al}. We use the fermionic representation of the 2DIM. Since the standard SCHA is
restricted to bosonic models, the new procedure is also an extension of the gaussian
approximation to fermionic 2D theories.  Our main result is an algebraic equation
which allows to get the behavior of the correlation length as function of $T-T_c$ and
$h$.

Let us stress that we are not introducing a new approximation but just a method to
determine its parameter. As it is well-known the SCHA is a non controlled
approximation, i.e. there is no perturbative parameter involved. It is then clear that
the same criticism can be made to the present proposal.

\vspace{1cm} We shall begin by depicting the main features of the standard SCHA. One
starts from a partition function

\begin{equation}
\Z\true=\int\D\mu e^{-S_{\text{true}}}
\end{equation}

\noindent where $\D\mu$ is a generic integration measure and $S_{\text{true}}$ is the
exact action. An elementary manipulation leads to

\begin{equation}\label{eq:Z}
\Z_{\text{true}}=\frac{\int\D\mu
e^{-(S_{\text{true}}-S_{\text{trial}})}\,e^{-S_{\text{trial}}}}{\int\D\mu
e^{-S_{\text{trial}}}}\int\D\mu e^{-S_{\text{trial}}}=\Z_{\text{trial}}\bra
e^{-(S_{\text{true}}-S_{\text{trial}})} \ket_{\text{trial}}
\end{equation}

\noindent for any trial action $S_{\text{trial}}$. Now, by means of the property

\begin{equation}
\bra e^{-f} \ket\geq e^{-\bra f \ket},
\end{equation}

\noindent for $f$ real, and taking natural logarithm in equation (\ref{eq:Z}), we
obtain Feynman's inequality \cite{Feynman}

\begin{equation}\label{Feynman}
\ln\Z\true\geq \ln\Z\trial - \bra S\true-S\trial\ \ket\trial.
\end{equation}
In general, $S\trial$ depends on some parameters, which are fixed by minimizing the
right hand side of the last equation.

At this point, in order to illustrate the procedure, we shall consider the well-known
SG model, with action

\begin{equation}
S\true=\int\difp\varphi(p)\frac{F(p)}{2}\varphi(-p) + \int d^2x\,\
\frac{\alpha}{\beta^2}\left[1-\cos(\beta\varphi)\right]
\end{equation}
where $\varphi(p)$ is a scalar field and $F(p)$ is usually of the form $F(p)\sim p^2$.
For simplicity, in this formula we have written the kinetic term in Fourier space but
we kept the interaction term in coordinate space.

As the trial action one proposes a quadratic one,

\begin{equation}
S\trial=\int\difp\left[\varphi(p)\frac{F(p)}{2}\varphi(-p) +
\frac{\Omega^2}{2}\varphi(p)\varphi(-p)\right],
\end{equation}
where $\Omega$ is the trial parameter. In order to perform the standard minimization
procedure, we first evaluate $\langle S\true-S\trial \rangle$. The result is
\begin{equation}\label{elegant}
\bra S\true-S\trial \ket\trial =
\V\left[\frac{\alpha}{\beta^2}\left(1-e^{-\unmedio\beta^2[I_1(\Omega)-I_1(\rho)]}\right)
-\frac{\Omega^2}{2}I_1(\Omega)\right].
\end{equation}

\noindent where
\begin{equation}
I_1(\Omega)=\int \frac{d^2p}{(2\pi)^2} \frac{1}{[F(p)+\Omega^2]},
\end{equation}
\noindent and $\rho$ is a normal-ordering parameter \cite{Coleman}.

Now inserting (\ref{elegant}) in equation (\ref{Feynman}), and extremizing the r.h.s.
with respect to $\Omega$, we finally obtain

\begin{equation}\label{gapscha}
\Omega^2-\alpha e^{-\beta^2/2 (I_1(\Omega)-I_1(\rho))}=0.
\end{equation}
This gap equation allows to extract a finite answer for $\Omega$, depending on the
mass parameter $\rho$ (the difference $I_1(\Omega)-I_1(\rho)$ is finite). Note that
the value of $\rho$ is completely arbitrary, if one chooses it to be equal to the
trial mass $\Omega$, the solution to the equation is

\begin{equation}\label{SGGaussian}
\Omega^2=\alpha .
\end{equation}
The same result is obtained if instead of $\rho=\Omega$ one takes
$\rho=\sqrt{\alpha}$.

\vspace{.5cm}

Let us now present an alternative route to determine $\Omega$. To this end we will
exploit a quantitative prediction of conformal invariance for 2D systems in the
scaling regime, away from the critical point. Starting from the so called 'c-theorem'
\cite{Zamolodchikov} Cardy \cite{Cardy} showed that the value of the conformal anomaly
$c$, which characterizes the model at the critical point, and the second moment of the
energy-density correlator in the scaling regime of the non-critical theory are related
by

\begin{equation} \label{Cardy}
\int d^2x\, |x|^2\, \bra\varepsilon(x)\varepsilon(0)\ket = \frac{c}{3\, \pi \,t^2\,(2
- \Delta_\varepsilon)^2},
\end{equation}
where $\varepsilon$ is the energy-density operator, $\Delta_\varepsilon$ is its
scaling dimension and $t\propto(T-T_c)$ is the coupling constant of the interaction
term that takes the system away from criticality. The validity of this formula has
been explicitly verified for several models \cite{Cardy} \cite{Cardy2}. For the SG
model, the energy density operator is given by the cosine term, its conformal
dimension is $\Delta_\varepsilon=\beta^2/4\pi$, $t$ is the coupling constant
$\alpha/\beta^2$ and the associated free bosonic CFT has $c=1$.

Now we claim that $\Omega$ can be determined in a completely different, not
variational way, by enforcing the validity of the above conformal identity for the
trial action. In other words, we will demand that the following equation holds:

\begin{equation} \label{Main}
\frac{\alpha^2}{\beta^4}\int d^2x\, |x|^2\, \bra
\cos\beta\varphi(x)\,\cos\beta\varphi(0) \ket\trial = \frac{1}{3\, \pi \,\,(2 -
\frac{\beta^2}{4\pi})^2},
\end{equation}

\noindent which is to be viewed as an equation for the mass parameter $\Omega$. Of
course, if one is interested in comparing the answer given by this formula with the
SCHA result, when evaluating the left hand side of (\ref{Main}) one must adopt a
regularizing prescription equivalent to the normal ordering implemented in the SCHA
calculation. A careful computation leads to the following gap equation:

\begin{equation}\label{gapcga}
(\frac{\Omega}{\rho})^{2(2-u)}=(\frac{\alpha}{\rho^2})^2\,\frac{3}{32}\,\frac{2-u}{u^2}
\end{equation}

\noindent where we have defined the variable $u=\beta^2/4\pi$ ($0\leq u <2$) and
$\rho$ is the normal ordering parameter, as before. We see that, as in the standard
SCHA equation (\ref{gapscha}), one has different answers for different choices of
$\rho$, but in this case, the results obtained for the values $\sqrt{\alpha}$ and
$\Omega$ are different. In any case one gets a non trivial dependence of $\Omega$ on
$\beta^2$ in contrast with the SCHA. This is interesting if one recalls the physical
meaning of mass gaps in the context of the SG model. Indeed, as it is well-known,
Dashen, Hasslacher and Neveu (DHN) \cite{DHN} have computed by semiclassical
techniques the mass spectrum for the SG model. It consists of a soliton (associated to
the fermion of the Thirring model) with mass

\begin{equation}\label{soliton}
M_{sol}=\frac{2-u}{\pi\,u}\,\sqrt{\alpha},
\end{equation}

\noindent and a sequence of doublet bound states with masses

\begin{equation}\label{doublets}
M_{N}=\frac{2(2-u)}{\pi\,u}\,\sin\left[N \frac{\pi\,u}{2(2-u)}\right]\, \sqrt{\alpha},
\end{equation}

\noindent with $N=1,2,...<(2-u)/u$. (From this last condition it is easy to see that
in order to have $N$ bound states one must have $u<2/(N+1)$. As a consequence there is
no bound state for $u>1$). More recently Zamolodchikov \cite{Zamolodchikov2}, by
reinterpreting Bethe ansatz results, has given exact expressions for this spectrum. In
particular for the soliton his formula coincides very well with (\ref{soliton}),
except for $u$ close to $2$, where it predicts a divergence. For simplicity here we
compare our results with equation (\ref{soliton}). The first thing to note is that the
masses in the SGM spectrum also depend on $u$ as our prediction given by equation
(\ref{gapcga}). Thus, in this respect our proposal seems to be able to improve the
standard gaussian prediction for the SGM, at least qualitatively. In order to perform
a more specific and quantitative discussion let us compare equations (\ref{gapcga})
and (\ref{soliton}) as functions of $u$. We set $\rho=\sqrt{\alpha}$, which
corresponds to the prescription employed by DHN when deriving (\ref{soliton}) and
(\ref{doublets}). The result is shown in Fig. 1 where one can observe a general
qualitative analogy between both curves. In particular, for $0.7\leq u \leq 1$ ($u=1$
corresponds to the free fermion point of the Thirring model and to the Luther-Emery
point in the backscattering model \cite{Luther-Emery}) our prediction is in full
agreement with the values of the soliton mass as computed by DHN. We want to stress
that for $u=1$ we get $\Omega/\sqrt{\alpha}=\sqrt{3/32}\approx0.30$ whereas the value
given by (\ref{soliton}) is $1/\pi\approx0.31$ (standard SCHA yields, of course,
$\Omega/\sqrt{\alpha}=1$).

\vspace{.5cm} Having checked the admissibility of our proposal in a model where exact
results are known, it is now desirable to explore a non-trivial problem. Let us
consider the 2D Ising model away from criticality ($T\neq T_c$ and $h\neq0$):
\begin{equation} \label{IsingAction}
S= S_0 + \int d^2x \left[t \, \epsilon(x)\,+ h \, \sigma(x)\right],
\end{equation}
\noindent where $S_0$ is the critical action, $t \propto(T-T_c)$, and $\epsilon(x)$
and $\sigma(x)$ are the energy-density and spin operators, respectively. We shall use
the fermionic representation for the above action. Thus $S_0$ is a free massless
Majorana action and $\epsilon\propto \PsiB \Psi$. On the other hand, the expression of
$\sigma(x)$ in terms of the Majorana fields is more involved. Indeed, by means of a
Jordan-Wigner transformation it can be written as an exponential of a fermionic
bilinear. In analogy to the usual SCHA method, we propose the following quadratic
trial action:
\begin{equation}
S_{trial}= S_0 + \Omega\, \int d^2x \, \epsilon(x),
\end{equation}
The conformal equation (\ref{Cardy}) for the present case takes the form
\begin{eqnarray} \label{Cardy2}
& \int d^2r\, r^2\,[t^2\,(2 - \Delta_\varepsilon)^2
\bra\varepsilon(r)\varepsilon(0)\ket\trial + h^2\,(2 - \Delta_\sigma)^2
\bra\sigma(r)\sigma(0)\ket\trial + \nonumber\\ & + 2\,t\,h\,(2 -
\Delta_\varepsilon)\,(2 - \Delta_\sigma) \bra\varepsilon(r)\sigma(0)\ket\trial]=
\frac{1}{6\, \pi},
\end{eqnarray}
\noindent where we have set $c=1/2$, which is the central charge corresponding to
Majorana free fermions and $\Delta_\varepsilon =1$ and $\Delta_\sigma=1/8$ are the
scaling dimensions of the corresponding operators. Now we have to evaluate the
v.e.v.'s in the trial theory. This will give us an equation for $\Omega$ as function
of t and h. The energy-energy and the energy-spin correlation functions have been
computed by Hecht \cite{Hecht} whereas the spin-spin correlator can be found in the
work of Wu, McCoy, Tracy and Barouch \cite{Wu}. As usual, one defines a correlation
length $\xi=1/4\Omega$ and considers the scaling limit given by
$\xi\rightarrow\infty$, $r\rightarrow\infty$, with $r/\xi$ fixed. The next step is to
use the expressions of the correlators for $(r/\xi)<<1$ and perform the corresponding
integrals. At this point we have to take into account that the correlation functions
are proportional to certain scaling functions $F_{\pm}(r/{\xi})$ where the $+$ and $-$
signs correspond to the cases $\Omega>0$ and $\Omega<0$ respectively. In other words,
the parameter $\Omega$ can be seen as defining a new "effective" critical temperature,
and the functions $F_{\pm}$ describe the scaling regime above and below this
temperature. Since we are approximating a magnetic perturbation of the system it is
clear that we must use the functions $F_{-}$. Thus we obtain the following equation
relating $\xi$, $h$ and $t$:

\begin{equation} \label{Main2}
t^2(4\xi)^2+C_1 h^2 (4\xi)^{15/4}+C2 t |h| (4\xi)^{23/8}=1
\end{equation}
where we have introduced the numerical constants $C_1=0.749661$ and $C_2=0.186966$.
The absolute value of the magnetic field in the second term comes from the fact that
$\bra\epsilon\sigma\ket\propto\bra\sigma\ket$ and the product $\bra\sigma\ket h$ has
to be positive since the magnetization and the magnetic field have the same
orientation. For $\xi$ fixed this equation gives a simple dependence of $h$ as
function of $t$. Indeed, for $h>0$ one has a slightly rotated semi ellipse in the
upper $h - t$ plane, and for $h<0$ one has its reflection over the $t=0$ axis.

An alternative form of equation (\ref{Main2}) is obtained if one introduces the
dimensionless combination $\chi={\mid h \mid}^{-\frac{8}{15}} /4\xi_0$:
\begin{eqnarray} \label{Main3}
\left(\frac{\xi}{\xi_0}\right)^2 + C_{1}
\,\chi^{\frac{-15}{4}}\,\left(\frac{\xi}{\xi_0}\right)^{\frac{15}{4}}\pm C_{2}
\,\chi^{\frac{-15}{8}}\,\left(\frac{\xi}{\xi_0}\right)^{\frac{23}{8}}=1.
\end{eqnarray}
\noindent The $+$ and $-$ signs in the third term of the left hand side correspond to
the cases $t>0$ and $t<0$, respectively. The action (\ref{IsingAction}) defines a one
parameter family of field theories which can be labelled by $\chi$. Moreover, the
particle content of the model is expected to undergo drastic changes as a function of
$\chi$ \cite{McCoy}.

In order to check the consistency of the above equations we first consider the limits
$h\rightarrow0$ and $t\rightarrow0$ separately. The first case corresponds to
$\chi\rightarrow\infty$ and one immediately obtains $\xi=\xi_0$, as expected. In the
second case one has $\chi\rightarrow0$ and then we get $\xi\sim{\mid h
\mid}^{-\frac{8}{15}}$, which is in agreement with the exact result obtained in
Ref.\cite{Z}. Let us mention that in this reference the constant of proportionality
was exactly determined to be 4.4, whereas our approximate computation yields 3.7.
Going back to the general case, we have solved equation (\ref{Main2}) numerically for
$\xi$ as function of $\chi$ for both $t>0$ and $t<0$. The results are plotted in Fig.
2. In the $t>0$ case the correlation length increases in a monotonous way from the
zero value and it approaches the $h=0$ value $\xi_0$ from below as
$\chi\rightarrow\Infinity$. In the $t<0$ case, although the behaviour of $\xi$ seems
very similar to the previous case, it presents a subtle difference shown in Fig. 3.
For $\chi\approx 2$ the correlation function goes over the $\xi_0$ value, reaches a
maximum and then tends to $\xi_0$ from above as $\chi\rightarrow\Infinity$. As this
behaviour depends on the value of the constants $C_1$ and $C_2$ we do not know whether
this is indeed a property of the Ising model or an artifact introduced by our
approximation.

 \vspace{1cm} To conclude, we have reconsidered the
well-known SCHA method in which a comparatively complex action is replaced by a
simpler quadratic system depending on a mass parameter $\Omega$ which is usually
determined through a variational calculation. Taking into account the
(1+1)-dimensional case, we have proposed an alternative way for evaluating $\Omega$.
Our proposal is based on a consequence of Zamolodchikov's \cite{Zamolodchikov}
c-theorem first derived by Cardy \cite{Cardy}. We have illustrated the idea by
considering the SG model. We showed that for this model our method gives a quite good
prediction for the behavior of the soliton mass as function of $\beta^2$ (see
equations (\ref{gapcga}) and (\ref{soliton}) and Fig. 1).

As a non-trivial application we have considered the 2D Ising model away from
criticality ($T\neq T_c$ and $h\neq0$). Starting from a continuum field theoretical
description in terms of Majorana fermions, we proposed a quadratic trial action
depending on a parameter $\Omega$ that defines an approximate correlation length
$\xi$. Our main result is given by equation (\ref{Main2})(or its alternative form
(\ref{Main3})) which allows to determine the parameter $\Omega$ (i.e. $\xi$) in terms
of the original physical parameters $t$ and $h$.

It would be interesting to test our approach in other models such as the continuum
version of the tricritical Ising model, which is described by the second model of the
unitary minimal series \cite{BPZ} \cite{FQS} with central charge $c=7/10$.
\vspace{0.5cm}

\begin{acknowledgments}

This work was supported by the Consejo Nacional de Investigaciones Cient\'{\i}ficas y
T\'ecnicas (CONICET) and Universidad Nacional de La Plata (UNLP), Argentina. We are
grateful to Bernard Jancovici for fruitful e-mail exchanges and for calling our
attention to ref.\cite{Zamolodchikov2}.
\end{acknowledgments}

\newpage

{\bf Figure caption}\\

Figure 1: Masses in units of $\sqrt{\alpha}$ as functions of $u$. The filled line is
$M_{sol}/\sqrt{\alpha}$, whereas the dashed line represents $\Omega/\sqrt{\alpha}$ as
given by equation \ref{gapcga}.\\

\vspace{2 cm}

Figure 2: Correlation length in units of $\xi_0$ as a function of $\chi$ for both
$t>0$ and $t<0$.

\vspace{2 cm}

Figure 3: As Fig. 2, showing details of the behaviour of $\xi(\chi)$ for $t>0$ and
$t<0$.
\end{document}